\documentclass[preprint,showpacs,preprintnumbers]{revtex4}
\usepackage{graphicx}
\usepackage{dcolumn}
\usepackage{bm}

\newcommand{\be}{\begin{equation}}
\newcommand{\en}{\end{equation}}
\newcommand{\bea}{\begin{eqnarray}}
\newcommand{\ena}{\end{eqnarray}}
\begin{document}

\title{Intermediate-Generalized Chaplygin Gas inflationary universe model }
\author{Ram\'on Herrera}
\email{ramon.herrera@ucv.cl}
\author{Marco Olivares}
\email{marco.olivares@ucv.cl}
\author{Nelson Videla}
\email{nelson.videla@ucv.cl}
\affiliation{Instituto de F\'{\i}sica, Pontificia Universidad Cat\'{o}lica de Valpara%
\'{\i}so, Avenida Brasil 2950, Casilla 4059, Valpara\'{\i}so, Chile.}
\date{\today}

\begin{abstract}
An intermediate inflationary universe model in the context of a
generalized Chaplygin gas is considered. For the matter we
consider two different  energy densities; a standard scalar field
and  a tachyon field, respectively. In general, we discuss the
conditions of an inflationary epoch for these models. We also, use
recent astronomical observations from Wilkinson Microwave
Anisotropy Probe seven year data for constraining the parameters
appearing in our models.
\end{abstract}

\maketitle




\section{Introduction}
The inflationary universe was introduced \cite{guth,infla} as a
manner  of addressing pressing problems (horizon, flatness,
monopoles, etc.) that were eating away at the bases   of the
otherwise rather prosperous Big-Bang model. The most significant
feature of the inflationary universe model is that it provides a
causal interpretation of the origin of the observed anisotropy of
the cosmic microwave background  radiation (CMB) and the structure
formation in the universe\cite{astro,astro2}.


Exact solutions exist for  power-law and de-Sitter inflationary
universes and they are created by exponential and constant scalar
potentials, see Ref.\cite{guth,power}. Exact solutions can also be
obtained for the scenario of intermediate inflation,  where the
scale factor, $a(t)$, increases as
\begin{equation}
a=\exp[\,A\,t^{f}],  \label{at}
\end{equation}
in which $A$ and $f$ are two constants; $A>0$ and $0<f<1$
\cite{Barrow1}. The expansion of this inflationary scenario  is
slower than  de-Sitter inflation, but faster than power law
inflation, this is the denotation why it is called "intermediate".
The intermediate model was originally formulated as an exact
solution, but it may be best inspired from the slow-roll
approximation.  From  the slow-roll approximation, it is possible
to have a spectrum of density perturbations which presents a
spectral index $n_s\sim 1$ and also in particular $n_s=1$
(Harrizon-Zel'dovich spectrum) for the value $f=2/3$
\cite{Barrow2}. However, the value  $n_s=1$ is disfavored  by the
current Wilkinson Microwave Anisotropy Probe (WMAP) observational
data\cite{astro,astro2}. Also, the tensor perturbations which
could be present in this model, through of the parametrized by the
tensor to scalar ratio $r$, which is significantly $r\neq
0$\cite{ratior,Barrow3}. On the other hand, the motivation to
study this expansion  becomes from string/M-theory, indicates that
in order to have a ghost-free action high order curvature
invariant corrections to the Einstein-Hilbert action must be
relative to the Gauss-Bonnet (GB) term{\cite{BD}}, where this
expansion appears to the low-energy string effective
action{\cite{KM,ART}} (see also, Ref.\cite{Varios1}).

On the other hand, the generalized Chaplygin gas (GCG) is other
aspirant for explaining the  acceleration of universe. The exotic
equation of state of the GCG is given by \cite{Chap}
\begin{equation}
p_{Ch}=-\frac{\alpha}{\rho_{Ch}^{\beta}}  \label{r1}
\end{equation}
where $\rho_{Ch}$ and $p_{Ch}$ are the energy density and pressure
of the GCG, $\beta$ is a constant in which $\beta\leq 1$, and
$\alpha$ is a positive constant. In particular, when $\beta$ = 1
corresponds to the original Chaplygin gas \cite{Chap}. Replacing,
Eq.(\ref{r1}) into the stress-energy conservation equation, the
energy density results
\begin{equation}
\rho_{Ch}=\left[\alpha+\frac{B}{a^{3(1+\beta)}}\right]
^{\frac{1}{1+\beta}}=\rho_{Ch0}\left[B_s+\frac{(1-B_s)}{a^{3(1+\beta)}}\right]^{\frac{1}{1+\beta}}.
\label{r2}
\end{equation}
Here, $a$ is the scale factor and B is a positive integration
constant. In this way, the GCG is characterized by two parameters,
$B_s=\alpha/\rho_{Ch0}^{1+\beta}$ and $\beta$. These parameter has
been confronted by observational data, see
Refs.\cite{const,const1}. In particular, the values of
$B_s=0.73_{-0.06}^{+0.06}$ and $\beta=-0.09_{-0.12}^{+0.15}$ was
obtained in Ref.\cite{const1}. Also, in Ref.\cite{27} the values
$0.81\lesssim B_s\lesssim 0.85$ and $0.2\lesssim\beta\lesssim 0.6$
were found from the observational data arising from Archeops for
the location of the first peak, BOOMERANG for the location of the
third peak, supernova and high-redshift observations. Recently,
the values of $B_s=0.775_{-0.0161-0.0338}^{+0.0161+0.037}$ and
$\beta=0.00126_{-0.00126-0.00126}^{+0.000970+0.00268}$ was
obtained from Markov Chain Monte Carlo method\cite{const2}.

The Chaplygin gas arises  as an effective fluid of a generalized
d-brane the space time, in a Born-Infeld action \cite{27} and
these models have been extensively analyzed  in Ref.\cite{Anali}.
In the model of Chaplygin inspired in an inflationary scenario
commonly the standard scalar field  drives inflation, in which the
energy density given by Eq.(\ref{r2}), can be extrapolate in the
Friedmann equation for archiving an appropriate inflationary
period \cite{Bertolami:2006zg}. However, also a tachyonic field in
a Chaplygin inflationary universe model was considered in
Ref.\cite{mon1}. The possibility of having Chaplygin models with
scalar field and tachyon field has been considered in
Ref.\cite{Bo}. The modification of the Friedmann equation is
realized from an extrapolation of Eq.(\ref{r2}), where we
identifying the density matter with the contributions of the
density energy associated to the standard scalar field or
tachyonic field \cite{27,mon1}. In this way, the GCG model may be
viewed as a variation  of gravity and there has been great
interest in the elaboration of early universe scenarios motivated
by string/M-theory\cite{mod}. It is well known that these
modifications can lead to significant changes in the early
universe.



In this paper  we would like to study intermediate-GCG inflationary
universe model in which  different types of energy densities  are
taken into account. In particular, (i) when the energy density is a
standard scalar field, and (ii) when the energy density is a tachyon
field. We will investigate the dynamic in both models and also we
shall utilize to the seven-year data WMAP to restrict the parameters
in our models. The outline of the paper is as follows. The next
section presents the dynamic of the intermediate-GCG Inflationary
scenario for our two models. Section \ref{sectpert} deals with the
calculations of cosmological perturbations. Finally, in
Sect.\ref{conclu} we conclude with our finding.

\section{Intermediate-GCG inflationary universe model}

It is well know that the GCG model can also be used to describe
the early universe. During inflation the gravity dynamics may give
rise to a modified Friedmann equation\cite{27}
\begin{equation}
H^2=\frac{\kappa}{3}\left(\alpha+\rho_\phi^{1+\beta}\right)^{\frac{1}{1+\beta%
}},  \label{a1}
\end{equation}
where $\kappa=8\pi/m_p^2$, $m_p$ is the reduced Planck mass,
$\rho_\phi$ is the energy density of the scalar field and
$H=\dot{a}/a$ is the Hubble parameter. This modification in the
Friedmann equation is the so-called Chaplygin inspired inflation
scenario\cite{27}. Following this idea, lately, some work has been
done, this involves Chaplygin inflationary universe model, see
Refs.\cite{mon1,mon2}.

In the following, we will considers two matter fields for
$\rho_\phi$; the standard scalar field and tachyonic field,
respectively. For convenience we use units in which $c=\hbar=1$.

\subsection{Standard scalar field}

We consider that the matter content of the universe is a standard scalar
field $\phi$, in which the energy density is given by $\rho_\phi=\frac{\dot{%
\phi}^2}{2}+V(\phi)$ and the pressure $p_\phi=\frac{\dot{\phi}^2}{2}-V(\phi)$
where $V(\phi)=V$ is the scalar potential. The conservation equation is
given by
\begin{equation}
\dot{\rho_\phi}+3\,H\,(\rho_\phi+p_\phi)=0,  \label{rho1}
\end{equation}
which is equivalent to the equation of motion of the standard scalar field
\begin{equation}
\ddot{\phi}+\,3H \; \dot{\phi}+V^{\prime }=0,  \label{phista}
\end{equation}
in which, $V^{\prime }=\partial V(\phi)/\partial\phi$ and the dots
mean derivatives with respect to time.

From Eqs.(\ref{a1}) and (\ref{rho1}), we get

\begin{equation}
\dot{\phi}^2=\frac{2}{\kappa} (-\dot{H})
\left[1-\alpha\left(\frac{\kappa}{3H^{2}}\right)^{1+\beta}\right]^{\frac{-\beta}{1+\beta}},
\label{fi}
\end{equation}
and the effective potential becomes
\begin{equation}
V=\frac{3}{\kappa}H^{2}
\left[1-\alpha\left(\frac{\kappa}{3H^{2}}\right)^{1+\beta}\right]^{\frac{1}{1+\beta}}+
\frac{1}{\kappa} \dot{H}
\left[1-\alpha\left(\frac{\kappa}{3H^{2}}\right)^{1+\beta}\right]^{\frac{-\beta}{1+\beta}}.
\label{pot}
\end{equation}
Note that for $\alpha=0$, the expression for $\dot{\phi}^2$ and
the scalar potential $V$ given by Eqs.(\ref{fi}) and (\ref{pot}),
reduced to typical
expression corresponding to standard inflation, where $\dot{\phi}^2=-2\dot{H}%
/\kappa$ and $V=(3H^2+\dot{H})/\kappa$ \cite{Barrow2}.

The solution for the standard scalar field $\phi$, using Eqs.(\ref{at}) and (%
\ref{fi}) is given by

\begin{equation}
\phi(t)-\phi_0=\frac{\mathcal{B}[t]}{K}\,,  \label{exf}
\end{equation}
where $\phi(t=0)=\phi_0$ is an integration constant, the constant $K\equiv%
\sqrt{6}(1+\beta)\left(\frac{\kappa}{3}\right)^{\frac{2-f}{4(1-f)}}\sqrt{%
(1-f)} (A f)^{\frac{-1}{2(1-f)}} \,\alpha^{\frac{1}{%
4(1-f)(1+\beta)}}$ and

\[
\mathcal{B}[t]\equiv B\left[\left(\frac{\kappa}{3}\right)^{1+\beta}\frac{%
\alpha\,t^{2(1-f)(1+\beta)}}{(A f)^{2(1+\beta)}};\frac{f}{2(1-f)(1+\beta)},%
\frac{2+\beta}{2(1+\beta)}\right].
\]
Here, $\mathcal{B}[t]$, is the incomplete Beta function\cite{20}
and without loss of generality $\phi_0=0$.

For the Hubble parameter $H(\phi)$, we get $
H(\phi)=Af(\mathcal{B}^{-1}[K\,\phi])^{f-1},$
 where $\mathcal{B}^{-1}$ represent the inverse function of the
incomplete Beta function.

In the slow-roll approximation, the first term of Eq.(\ref{pot})
dominate the effective potential at large value of $\phi$ and
using Eqs.(\ref{pot}) and (\ref{exf}), we have

\begin{equation}
V(\phi)=\left[\left(\frac{3A^{2}f^{2}}{\kappa}\right)^{1+\beta}(\mathcal{B}%
^{-1}[K\,\phi])^{-2(1-f)(1+\beta)} -\alpha\right]^{\frac{1}{1+\beta}}.
\label{potapp}
\end{equation}
Note that we would have obtained the same potential $V(\phi)$ represented by
Eq.(\ref{potapp}), considering the set of slow-roll conditions, where $\dot{%
\phi}^2 \ll V(\phi)$ and $\ddot{\phi}\ll 3H\dot{\phi}$.

The dimensionless slow-roll parameters in this case become
$
\varepsilon\equiv-\frac{\dot{H}}{H^2}=\frac{1-f}{Af(\mathcal{B}%
^{-1}[K\,\phi])^{f}} , $ and
$
\eta\equiv-\frac{\ddot{H}}{H \dot{H}}=\frac{2-f}{Af(\mathcal{B}%
^{-1}[K\,\phi])^{f}}. $ The inflationary scenario takes place when
the slow-roll parameter $\varepsilon<1$ or analogously when
$\ddot{a}>0$. Therefore, the
condition for inflation to occur is satisfied when the standard field $\phi>%
\frac{1}{K}\,\mathcal{B}\left[\left(\frac{1-f}{Af}\right)^{1/f}\right]. $

Using Eq.(\ref{exf}), the number of e-folds $N$ between two values
of cosmological times $t_1$ and $t_2$ or analogously between two
different values of $\phi$, in which $\phi(t=t_1)=\phi_1$ and
$\phi(t=t_2)=\phi_2$, becomes
\begin{equation}
N=\int_{t_1}^{t_{2}}\,H\,dt=A\,\left[(t_{2})^{f}-(t_{1})^{f}\right] =A\,%
\left[(\mathcal{B}^{-1}[K\,\phi_{2}])^{f}-(\mathcal{B}^{-1}[K\,\phi_{1}])^{f}%
\right].  \label{N}
\end{equation}

Considering that the inflationary scenario begins at the earliest
possible scenario in which $\varepsilon=1$ \cite{Barrow3}, then
the
scalar field $\phi_1$, is given by $\phi_{1}=\frac{1}{K}\,\mathcal{B}\left[%
\left(\frac{1-f}{Af}\right)^{1/f}\right].$

\subsection{Tachyon field}

For the case of the tachyonic field, the energy density and the pressure are
given by $\rho _{\phi }=\frac{V(\phi )}{\sqrt{1-\dot{\phi}^{2}}}$ and $%
P_{\phi }=-V(\phi )\sqrt{1-\dot{\phi}^{2}}$, respectively. Here,
$\phi $ represents the tachyon field and $V(\phi )=V$ is the
tachyonic potential. The equation of motion for the tachyonic
field from Eq.(\ref{rho1}), is given by
\begin{equation}
\frac{\ddot{\phi}}{1-\dot{\phi}^{2}}\,+3H\;\dot{\phi}+\frac{V^{\prime }}{V}%
=0.  \label{Tkey_01}
\end{equation}

Using, Eqs.(\ref{a1}) and (\ref{Tkey_01}), we get
\begin{equation}
\dot{\phi}^{2}=-\frac{2\dot{H}}{3H^{2}}\left[  1-\alpha\left(
\frac{\kappa
}{3H^{2}}\right)  ^{1+\beta}\right]  ^{-1}, \label{Tfi}%
\end{equation}
and the tachyonic potential as function of the Hubble parameter
$H$ and $\dot{H}$, becomes
\begin{equation}
V=\left[  \left(  \frac{3}{\kappa}\right)  ^{1+\beta}H^{2(1+\beta)}%
-\alpha\right]
^{\frac{1}{1+\beta}}\sqrt{1+\frac{2\dot{H}}{3H^{2}}\left[
1-\alpha\left(  \frac{\kappa}{3H^{2}}\right)  ^{1+\beta}\right]
^{-1}}.
\label{Tpot}%
\end{equation}
Again, when $\alpha=0$ the expressions for the velocity of the
tachyonic field $\dot{\phi}$ and $V$ reduced to standard tachyonic
model, where $\dot{\phi }=\sqrt{-2\dot{H}/(3H^{2})}$ and
$V=(3/\kappa)H^{2}\sqrt{1+2\dot{H}/(3H^{2})}$ (see
Ref.\cite{mon3}).

From Eqs.(\ref{at}) and (\ref{Tfi}), the solution for the tachyonic field
becomes
\begin{equation}
\phi(t)-\phi_0=\frac{\widetilde{\mathcal{B}}[t]}{\widetilde{K}}\text{ ,}
\label{Texf}
\end{equation}%
where the constant $\widetilde{K}\equiv \frac{\sqrt{6}(1+\beta )\sqrt{1-f}(%
\frac{\kappa }{3})^{\frac{2-f}{4(1-f)}}\alpha ^{\frac{2-f}{4(1+\beta )(1-f)}}%
}{(Af)^{\frac{1}{2(1-f)}}}$, and
\[
\widetilde{\mathcal{B}}[t]\equiv B\left[\left( \frac{\kappa
}{3}\right)
^{1+\beta }\frac{\alpha t^{2(1+\beta )(1-f)}}{(Af)^{2(1+\beta )}};\frac{2-f}{%
4(1+\beta )(1-f)},\frac{1}{2}\right].
\]
Here, again $\widetilde{\mathcal{B}}$ is the incomplete Beta function.
Now, by using Eqs.(\ref%
{at}) and (\ref{Texf}), the Hubble parameter as a function of the
tachyon field, becomes $H(\phi
)=Af(\widetilde{\mathcal{B}}^{-1}[\widetilde{K}\phi ])^{f-1}$, where
$\widetilde{\mathcal{B}}^{-1}$ represent the inverse function of the
incomplete Beta function and $\phi_0=0$.

Analogously, as the case of the standard scalar field, during the
slow-roll approximation, the first term of Eq.(\ref{Tpot}) dominate
the effective potential at large value of $\phi$ and from
Eqs.(\ref{at}) and (\ref{Texf}), we get

\begin{equation}
V(\phi )= \left[ \left( \frac{3}{\kappa }\right) ^{1+\beta }\frac{\left(
Af\right) ^{2(1+\beta )}}{\left( (\widetilde{\mathcal{B}}^{-1}[\widetilde{K}%
\phi ])^{1-f}\right) ^{2(1+\beta )}}-\alpha \right] ^{\frac{1}{1+\beta }}.
\label{SRVF}
\end{equation}

In addition, note that again we would have obtained the same
tachyonic potential,
considering the set of slow-roll conditions for the tachyonic field, where $%
\dot{\phi}^2 \ll 1$ and $\ddot{\phi}\ll 3H\dot{\phi}$.

Again, as before now we can write the dimensionless slow-roll
parameters for the case of the  tachyonic field. Considering
Eqs.(\ref{at}) and (\ref{Texf}), we get $ \varepsilon
=\frac{1-f}{Af}(\widetilde{\mathcal{B}}^{-1}[\widetilde{K}\phi
])^{-f}$, and\, $\eta
=\frac{2-f}{Af}(\widetilde{\mathcal{B}}^{-1}[\widetilde{K}\phi
])^{-f}\label{Tep2}. $

The number of e-folds between times $t_{1}$ and $t_{2}$ using Eq.(\ref{Texf}%
) is given by
\begin{equation}
N=\int_{t_{1}}^{t_{2}}\,H\,dt=A\,\left[ (t_{2})^{f}-(t_{1})^{f}\right] =A\,%
\left[ (\widetilde{\mathcal{B}}^{-1}[\widetilde{K}\phi _{2}])^{f}-(%
\widetilde{\mathcal{B}}^{-1}[\widetilde{K}\phi _{1}])^{f}\right] .
\label{NOEF}
\end{equation}

Analogously, as the case of the standard field, the inflation
begins
at the earliest possible scenario, in which $\phi_{1}=\frac{1}{\widetilde{K}%
}\,\widetilde{\mathcal{B}}\left[  \left(  \frac{1-f}{Af}\right)
^{1/f}\right]  .$

\section{Cosmological Perturbations\label{sectpert}}

In this section we will analyze the scalar and tensor perturbations for our
models, where the matter content of the universe are the standard scalar
field and the tachyonic field, respectively.

\subsection{Standard scalar field}
In the following, we will consider  the power spectra of scalar
and tensor perturbations to the metric in Chaplygin inflation. We
introduce the gauge invariant quantity\cite{Bardee,Malik:2008im}
$$
\zeta=H\,+\frac{\delta\rho}{\dot{\rho}},
$$
where $\psi$ is the gravitational potential. On slices of uniform
density $\zeta$ reduces to the curvature perturbation. A
fundamental characteristic  attribute of $\zeta$ is that it is
nearly constant on super-horizon scales. This feature, result to
be a consequence of stress-energy conservation and  does not
depend on the gravitational dynamics\cite{Bassett:2005xm} (see
also, Ref.\cite{wand2}). In this context, it continues unchanged
in Chaplygin inflation\cite{Bertolami:2006zg,Zarrouki:2010jt}. In
this form, the power spectrum
related to curvature spectrum, could be written as ${\mathcal{%
P}_{\mathcal{R}}}\simeq\langle\zeta^2\rangle$. It can be shown
that on super-horizon scales, the curvature perturbation on slices
of uniform density is equivalent  to the comoving curvature
perturbation. Therefore, for the spatially flat gauge, we have
$\zeta=H\frac{\delta\phi}{\dot{\phi}}$, in which
$|\delta\phi|=H/2\pi$ \cite%
{Liddle}.

In this way, the power spectrum considering  Eq.(\ref{fi}), is
given by
\begin{equation}
{\mathcal{P}_{\mathcal{R}}}\simeq\frac{\kappa}{8\pi^{2}}
 H^{4}(-\dot{H})^{-1}
 \left[1-\alpha\left(\frac{\kappa}{3H^{2}}\right)^{1+\beta}\right]^{\frac{\beta}{1+\beta}},
\label{dp}
\end{equation}
or equivalently in terms of the standard scalar field $\phi$
\begin{equation}
{\mathcal{P}_{\mathcal{R}}}\simeq\frac{\kappa}{8\pi^{2}}\frac{(Af)^{3}}{1-f}
(\mathcal{B}^{-1}[K\,\phi])^{-(2-3f)}\left[1-\alpha\left(\frac{\kappa}
{3A^{2}f^{2}}\right)^{1+\beta}
(\mathcal{B}^{-1}[K\,\phi])^{2(1-f)(1+\beta)}\right]^{\frac{\beta}{1+\beta}}.
\label{prs}
\end{equation}

The power spectrum ${\mathcal{P}_{\mathcal{R}}}$, also can be expressed in
terms of the number of e-folds $N$, as

\begin{equation}
{\mathcal{P}_{\mathcal{R}}}=\frac{\kappa}{8\pi^{2}}\frac{(Af)^{3}}{1-f}
\left[\frac{Af}{1+f(N-1)}\right]^{\frac{2-3f}{f}}\left[1-\alpha\left(\frac{\kappa}
{3A^{2}f^{2}}\right)^{1+\beta}
\left[\frac{1+f(N-1)}{Af}\right]^{\frac{2(1-f)(1+\beta)}{f}}\right]^{\frac{\beta}{1+\beta}}.\label{Ns}
\end{equation}

Numerically from Eq.(\ref{Ns}) we obtained  a constraint for the
parameter $A$. In fact, we can obtain the value of the parameter
$A$ for  given values of $f$, $\alpha$ and $\beta$ parameters when
number $N$ and the  power spectrum ${\mathcal{P}_{\mathcal{R}}}$
are given. In particular, for the values
${\mathcal{P}_{\mathcal{R}}}=2.4\times 10^{-9}$, $N=60$, $f=1/2$
and $\kappa=1$, we obtained that for the pair ($\alpha=0.775$,
$\beta=0.00126$)\cite{const2}, which corresponds to the parameter
$A\simeq 8.225\times 10^{-2}$, for the pair ($\alpha=0.81$,
$\beta=0.2$)\cite{27}, corresponds to $A\simeq 2.635\times
10^{-2}$ and for the pair ($\alpha=0.85$, $\beta=0.6$)\cite{27},
which corresponds to $A\simeq 8.407\times 10^{-5}$.

On the other hand, the scalar spectral index $n_s$ is given by the expression $n_s=d\ln{%
\mathcal{P}_{\mathcal{R}}}/\ln k$ and considering Eq.(\ref{prs}),
we get
\begin{equation}
n_s \simeq
1-\frac{2-3f}{Af(\mathcal{B}^{-1}[K\,\phi])^{f}}-\left[\frac{\alpha_0\beta\gamma}{Af(1+\beta)}\right]
\left[1-\alpha_0(\mathcal{B}^{-1}[K\,\phi])^{\gamma}\right]^{-1}(\mathcal{B}^{-1}[K\,\phi])^{\gamma-f},
\label{Tnsa}
\end{equation}
where the constants
$\alpha_0=\alpha\left(\frac{\kappa}{3A^2f^2}\right)^{1+\beta}$ and
$\gamma=2(1-f)(1+\beta)$, respectively.

From Eq.(\ref{Tnsa}), we clearly see that $n_s\neq 1$,  for
$f=2/3$ (recall that $1>f>0$). However, as occurs in
Ref.\cite{Barrow3}, $n_s=1$ for the value $f=2/3$, where the scale
factor increases as $a(t)\sim e^{t^{2/3}}$. Also, we noted that in
the limit $\alpha\rightarrow 0$, the scalar spectral index $n_s$,
given by Eq.(\ref{Tnsa}), coincides with that corresponding to
intermediate-inflationary model, where $n_s=1-C_1/\phi^2$ with $%
C_1=8(1-f)(2-3f)/f^2$, see Ref.\cite{Barrow3}.

The scalar spectral index $n_s$ in terms of the number of e-folds $N$,
becomes
\begin{equation}
n_s \simeq
1-\frac{2-3f}{1+f(N-1)}-\left[\frac{\alpha_0\beta\gamma}{Af(1+\beta)}\right]
\left[1-\alpha_0([1+f(N-1)]/Af)^{\gamma/f}\right]^{-1}\left[\frac{1+f(N-1)}{A\,f}\right]^{(\gamma-f)/f}.
\label{Tnsa2}
\end{equation}

On the other hand, the generation of tensor perturbations during
the scenario inflationary would produce gravitational wave
\cite{Bassett:2005xm}. The corresponding spectrum is
\begin{equation}
{\mathcal{P}}_g =\,8\,\kappa\,\left(\frac{H}{2\pi}\right)^2= \frac{2\kappa}{%
\pi^{2}}A^{2}f^{2}(\mathcal{B}^{-1}[K\,\phi])^{-2(1-f)}.  \label{ag}
\end{equation}

\begin{figure}[th]
\includegraphics[width=3.3in,angle=0,clip=true]{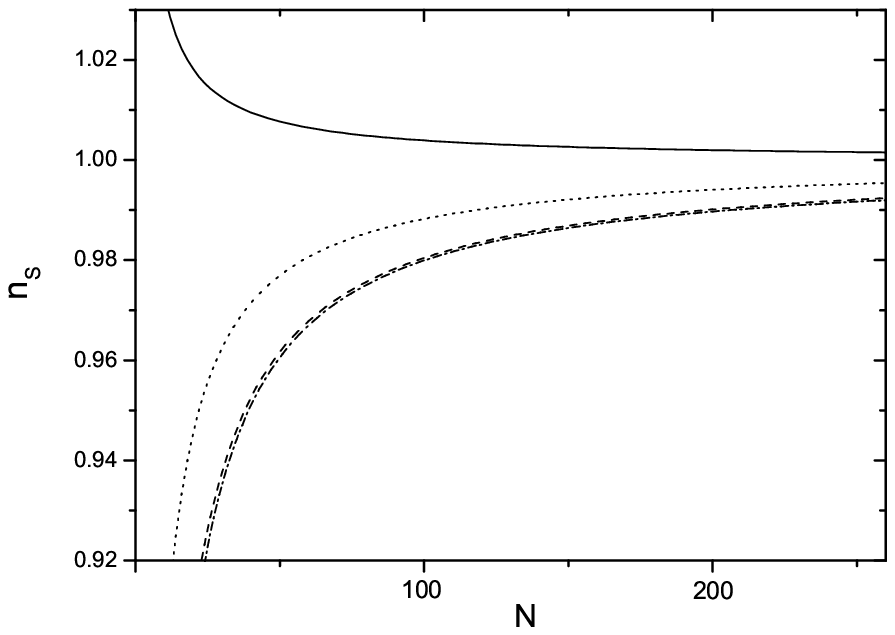}
\includegraphics[width=3.5in,angle=0,clip=true]{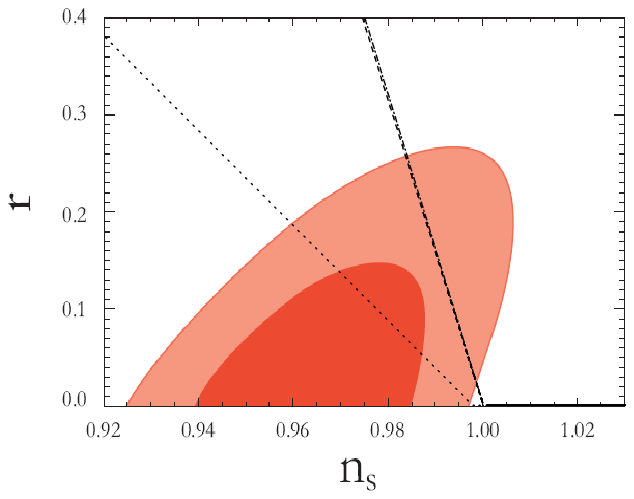}
\caption{ The upper panel shows the evolution of the scalar
spectrum index $n_s$ versus the number of e-folds $N$. The lower
panel shows the contour plot for the parameter $r$  as a function
of the $n_s$ at lowest order, for the case of the standard field.
Here, from WMAP seven-years data\cite{astro}, two-dimensional
marginalized constraints (68$\%$ and 95$\%$ confidence levels) on
inflationary parameters $r$ and $n_s$. Dotted, dashed, solid and
dot-dashed lines are for the pairs ($\alpha=0.81$, $\beta=0.2$),
($\alpha=0.775$, $\beta=0.00126$), ($\alpha=0.85$, $\beta=0.6$),
and the standard intermediate model ($\alpha=0$), respectively. In
both panels we have taken the values $\rho_{Ch0}=1$, $f=1/2$,
$\kappa=1$ and $A\simeq 2.635\times 10^{-2}; 8.225\times 10^{-2};
8.407\times 10^{-5}$, respectively.
 \label{A}}
\end{figure}
An important observational quantity is the tensor to scalar ratio $r$, which
is defined as $r=\left(\frac{{\mathcal{P}}_g}{P_{\mathcal{R}}}\right)$. From
Eqs.(\ref{prs}) and (\ref{ag}) we write the tensor to scalar ratio as
\begin{equation}
r(\phi)\simeq\frac{16(1-f)}{Af(\mathcal{B}^{-1}[K\,\phi])^{f}}
\left[1-\alpha\left(\frac{\kappa} {3A^{2}f^{2}}\right)^{1+\beta}
(\mathcal{B}^{-1}[K\,\phi])^{2(1-f)(1+\beta)}\right]^{\frac{-\beta}{1+\beta}}.
\label{Rk}
\end{equation}

Combining  Eqs.(\ref{N}) and (\ref{Rk}), we can write the
tensor-scalar ratio $r$ in terms of the number $N$, as
\begin{equation}
r(N)\simeq\frac{16(1-f)}{1+f(N-1)}\left[1-\alpha\left(\frac{\kappa}
{3A^{2}f^{2}}\right)^{1+\beta}
\left[\frac{1+f(N-1)}{Af}\right]^{\frac{2(1-f)(1+\beta)}{f}}\right]^{\frac{-\beta}{1+\beta}}.
\end{equation}

In Fig.(\ref{A}), the upper panel shows the evolution of the
scalar spectrum index $n_s$ versus the number of e-folds $N$, and
the lower panel shows the contour plot for the parameter $r$  as a
function of the $n_s$ at lowest order, for different values of the
parameters-GCG, $\alpha$ and $\beta$ in the case of the standard
scalar field. In particular, the dotted, dashed, solid and
dot-dashed lines are for the pairs ($\alpha=0.81$, $\beta=0.2$)
see Ref.\cite{27}, ($\alpha=0.775$, $\beta=0.00126$)\cite{const2},
($\alpha=0.85$, $\beta=0.6$)\cite{27}, and the standard
intermediate model ($\alpha=0$)\cite{Barrow3}, respectively. Here,
we have used the value  $\rho_{Ch0}=1$, then the parameter
$B_s=\alpha/\rho_{Ch0}^{1+\beta}=\alpha$. From the upper panel, we
noted that the $n_s$ graphs for the pair ($\alpha=0.775$,
$\beta=0.00126$) present a small displacement with respect to the
number of e-folds $N$, when compared to the results obtained in
the standard intermediate model, in which $\alpha=0$.

On the other hand, from Ref.\cite{astro}, two-dimensional
marginalized constraints (68$\%$ and 95$\%$ confidence levels) on
inflationary parameters $r$ and $n_s$, the spectral index of
fluctuations, defined at $k_0$ = 0.002 Mpc$^{-1}$. In order to
write down values that relate the tensor to scalar ratio and the
spectral index we numerically solved Eqs. (\ref{Tnsa}) and
(\ref{Rk}). Also, we have used the values  $f=1/2$, $\kappa=1$ and
for the parameter $A$ the values $A\simeq 2.635\times 10^{-2};
8.225\times 10^{-2}; 8.407\times 10^{-5}$, respectively. We noted
that the pairs ($\alpha=0.81$, $\beta=0.2$) and ($\alpha=0.775$,
$\beta=0.00126$), the model is well supported by the data as could
be seen from Fig.(\ref{A}). Also, we noted that the pair
($\alpha=0.85$, $\beta=0.6$) given by solid line, becomes
disfavored from observational data, since the spectral index
$n_s>1$. Also, we noted that for this pair $r\sim 0$  (solid
line). We have found that the pair ($\alpha=0.775$,
$\beta=0.00126$), present a small displacement
 in relation to the standard intermediate model that corresponds to $\alpha=0$,
 as could be seen from  the Fig.(\ref{A}).

In this way, we have shown that the  intermediate-GCG inflationary
model  is less restricted than analogous ones standard intermediate
inflationary models due to the introduction of new parameters, i.e.,
$\alpha$ and $\beta$ parameters.

\subsection{Tachyon field}

For a tachyonic field the power spectrum of the curvature
perturbations is given by ${\mathcal{P}_{R}}=\left(
\frac{H^{2}}{2\pi \dot{\phi}}\right)  ^{2}\frac{1}{Z_{S}}$
\cite{Tper}, where $Z_{S}=V(1-\dot{\phi }^{2})^{-3/2}\approx
V$\cite{Tper2}.  Following Ref.\cite{Tper2}, the power spectrum
${\mathcal{P}_{R}}$ is approximated to be
${\mathcal{P}_{R}}\simeq\left( \frac{H^{2}}{2\pi\dot{\phi}}\right)
^{2}\frac{1}{V}$. From Eq.(\ref{Tfi}) and considering
Eq.(\ref{Texf}), we write the power spectrum in terms of the
tachyonic field in the following way
\begin{equation}
{\mathcal{P}_{\mathcal{R}}}\simeq\frac{\kappa}{8\pi^{2}}\frac{\left(
Af\right)  ^{3}}{(1-f)}(\widetilde{\mathcal{B}}^{-1}[\widetilde{K}%
\phi])^{3f-2}\left[  1-\;\alpha\left(
\frac{\kappa}{3(Af)^{2}}\right)
^{1+\beta}(\widetilde{\mathcal{B}}^{-1}[\widetilde{K}\phi])^{2(1+\beta
)(1-f)}\right]  ^{\frac{\beta}{1+\beta}}. \label{PST}%
\end{equation}
The scalar spectral index $n_{s}$, using Eq.(\ref{Texf}), is given
by

\begin{equation}
n_{s}\simeq
1-\frac{2-3f}{Af(\widetilde{\mathcal{B}}^{-1}[\widetilde{K}\,\phi])^{f}}-
\left[\frac{\alpha_0\beta\gamma}{Af(1+\beta)}\right]
\left[1-\alpha_0(\widetilde{\mathcal{B}}^{-1}[\widetilde{K}\,\phi])^{\gamma}\right]^{-1}
(\widetilde{\mathcal{B}}^{-1}[\widetilde{K}\,\phi])^{\gamma-f}. \label{NSF}%
\end{equation}
Again, as the case of the standard scalar field from
Eq.(\ref{NSF}), we  see that $n_s\neq 1$,  for the case $f=2/3$.


On the other hand, the amplitude of tensor perturbations
${\mathcal{P}}_{g}$, is given by
\begin{equation}
{\mathcal{P}}_{g}=\,8\,\kappa\,\left(  \frac{H}{2\pi}\right)  ^{2}%
=\frac{2\kappa}{\pi^{2}}A^{2}f^{2}(\widetilde{\mathcal{B}}^{-1}[\widetilde
{K}\phi])^{-2(1-f)}. \label{Tag}%
\end{equation}

From expressions (\ref{PST}) and (\ref{Tag}) we write the tensor
to scalar ratio as
\begin{equation}
r(\phi)=\frac{16(1-f)}{Af}(\widetilde{\mathcal{B}}^{-1}[\widetilde{K}%
\phi])^{-f}\left[  1-\alpha\left(  \frac{\kappa}{3(Af)^{2}}\right)
^{1+\beta
}(\widetilde{\mathcal{B}}^{-1}[\widetilde{K}\phi])^{2(1+\beta)(1-f)}\right]
^{-\frac{\beta}{1+\beta}}. \label{TRk2}%
\end{equation}
Again, we noted that when $\alpha\rightarrow 0$ and considering
Eqs.(\ref{NSF}) and (\ref{TRk2}) the consistency relations at
lowest order, $n_s=n_s(r)$, reduced to standard tachyonic model,
where $n_s=1-\frac{2-3f}{16(1-f)}\,r$, see Ref.\cite{mon3}.

We noted numerically from Eqs.(\ref{NSF}) and (\ref{TRk2}) that
the trajectories in the $n_s -r$ plane between standard field and
tachyon field can not be distinguished at lowest order. This
coincidence in the consistency relations,  $n_s = n_s(r)$ between
standard field and tachyon field,  has already been noted in
Ref.\cite{igual}. Nevertheless, the tachyon field  inflationary
leads to a deviation at second order in the consistency relations,
where the spectral index at second order $n_s^{(2)}$,
becomes\cite{igual}
\begin{equation}
n_s^{(2)}\approx-(2\varepsilon^2+2[2C_1+3-2C_2]\varepsilon\eta+2C_1\eta\gamma),\label{nsT}
\end{equation}
where the product
$\eta\gamma=(9m_p^4/2)[2V''V'/V^4-10V''V'^2/V^5+9V'^4/V^6]$, the
constant $C_1$ is a numerical constant approximately
$C_1\simeq-0.72$ and the constant $C_2$; is $C_2=0$ in the case of
the standard scalar field and $C_2=1/6$ for tachyon field,
respectively. Following Ref.\cite{igual}, the expression for the
tensor to scalar ratio at second order $r^{(2)}$, in the tachyon
field is given by
\begin{equation}
r^{(2)}\approx16\varepsilon(2C_1\eta-2C_2\varepsilon).\label{rsT}
\end{equation}
\begin{figure}[th]
\includegraphics[width=5.0in,angle=0,clip=true]{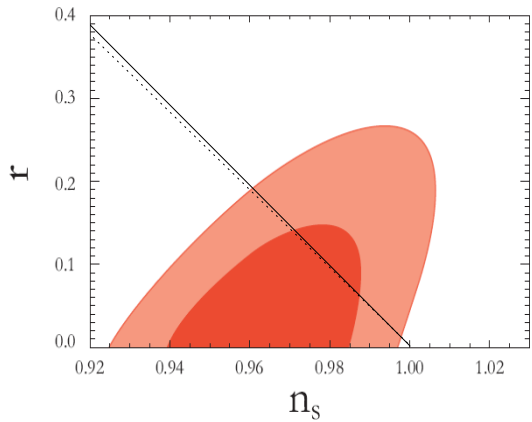}
\caption{Contour plot for the parameter $r$  as a function of the
$n_s$ for the pair ($\alpha=0.81$, $\beta=0.2$) in the case of the
tachyonic  field. Solid and  dotted  lines are for the
trajectories at lowest order and at second order, respectively.
Again, as before in drawing the graphs we took $A\simeq
2.635\times 10^{-2}$, $\rho_{Ch0}=1$, $f=1/2$ and $\kappa=1$.
 \label{B}}
\end{figure}
%

In Fig.(\ref{B}), we show the dependence of the tensor to scalar
ratio $r$ on the spectral index $n_s$, for the pair
($\alpha=0.81$, $\beta=0.2$) in the case of the tachyonic field.
Solid and dotted  lines are for the trajectories at lowest order
and at second order, respectively.  In order to write down values
that relate the tensor to scalar ratio and the spectral index, we
numerically solved Eqs.(\ref{NSF}), (\ref{TRk2}), (\ref{nsT}) and
(\ref{rsT}). Again as before, we have used the values $A\simeq
2.635\times 10^{-2}$, $\rho_{Cho}=1$, $f=1/2$ and $\kappa=1$. We
observed numerically, that the trajectories in the $n_s -r$ plane
for the tachyonic field, when we used the second-order corrections
to our analysis at first-order in slow roll, are small and this
correction can be neglected to a very good approximation.

\section{Conclusions \label{conclu}}

In this paper we have investigated the intermediate inflationary
model in GCG. In the slow-roll approximation we have found
solutions of the Friedmann equations for a flat universe
containing  a standard scalar field or a tachyonic field,
respectively. In particular, for both scalar fields and from the
scenario of intermediate inflation,  we have  obtained explicit
expressions for the corresponding, effective potential, power
spectrum of the curvature perturbations, tensor to scalar ratio
and scalar spectrum index.

For the scalar field, we have considered the constraints on the
parameters of the GCG, from the WMAP seven year data. Here, we
have taken the constraint $r-n_s$ plane at lowest order in the
slow roll approximation. In order to write down values that relate
the tensor to scalar ratio and the spectral index we numerically
solved Eqs. (\ref{Tnsa}) and (\ref{Rk}). We noted that the pairs
($\alpha=0.81$, $\beta=0.2$) and ($\alpha=0.775$,
$\beta=0.00126$), the model is well supported by the data as could
be seen from  Fig.(\ref{A}). Also, we noted that the pair
($\alpha=0.85$, $\beta=0.6$) given by solid line, becomes
disfavored from observational data, since the spectral index
$n_s>1$. We have found that the pair ($\alpha=0.775$,
$\beta=0.00126$), present a small displacement in relation to the
standard intermediate model that corresponds to $\alpha=0$, as
could be seen from the  Fig.(\ref{A}). In particular, we have used
the values $\rho_{Ch0}=1$, $f=1/2$, $\kappa=1$ and $A\simeq
2.635\times 10^{-2}; 8.225\times 10^{-2}; 8.407\times 10^{-5}$,
respectively.

For the tachyonic field, we noted numerically from Eqs.(\ref{NSF})
and (\ref{TRk2}) that the trajectories in the $n_s -r$ plane
between standard field and tachyon field can not be distinguished
at lowest order. However, we have obtained the dependence of the
tensor to scalar ratio $r$ on the spectral index $n_s$ at second
order. In order to write down values that relate the tensor to
scalar ratio and the spectral index at second order, we
numerically solved Eqs.(\ref{NSF}), (\ref{TRk2}), (\ref{nsT}) and
(\ref{rsT}), for the pair ($\alpha=0.81$, $\beta=0.2$). In this
case, we observed numerically that the trajectories in the $n_s
-r$ plane, when we used the second-order corrections to our
analysis with respect to the first-order corrections in slow roll
are small, as can be seen from Fig.(\ref{B}).

Finally, we have shown that the  intermediate-GCG inflationary
models are less restricted than analogous ones standard
intermediate inflationary models due to the introduction of new
parameters, i.e., $\alpha$ and $\beta$ parameters. The
incorporation of these parameters gives us a freedom that allows
us to modify the standard intermediate model by simply modifying
the corresponding values of the parameters $\alpha$ and $\beta$.

\begin{acknowledgments}
R.H. was supported by COMISION NACIONAL DE CIENCIAS Y TECNOLOGIA
through FONDECYT grants  N$^0$ 1090613,  N$^0$ 1110230 and by
DI-PUCV grant 123.703/2009. M.O. was supported by Proyecto D.I.
PostDoctorado 2012 PUCV. N.V. was supported by Proyecto
Beca-Doctoral CONICYT N$^0$ 21100261.
\end{acknowledgments}


\end{document}